\documentclass[11pt]{article}
\usepackage{moriond,epsfig}
\usepackage{graphicx}
\usepackage{amstext}

\def\be{\begin{equation}}
\def\ee{\end{equation}}
\def\bea{\begin{eqnarray}}
\def\eea{\end{eqnarray}}

\begin{document}
\vspace*{4cm}
\title{A MULTISCALE REGULARIZED RESTORATION ALGORITHM FOR XMM-NEWTON DATA}

\author{H. Bourdin, E. Slezak, A. Bijaoui}

\address{Observatoire de la C\^ote d'Azur, B.P. 4229, F-06304 NICE CEDEX 4, France \\ bourdin@obs-nice.fr, slezak@obs-nice.fr, bijaoui@obs-nice.fr}

\author{M. Arnaud}
\address{Service d'Astrophysique, CEA Saclay, F-91191 Gif-sur-Yvette, France \\ marnaud@discovery.saclay.cea.fr}

\maketitle

\abstracts{We introduce a new multiscale restoration algorithm for images with few photons counts and its use for denoising XMM data. We use a thresholding of the wavelet space so as to remove the noise contribution at each scale while preserving the multiscale information of the signal. Contrary to other algorithms the signal restoration process is the same whatever the signal to noise ratio is. Thresholds according to a Poisson noise process are indeed computed analytically at each scale thanks to the use of the unnormalized Haar wavelet transform. Promising preliminary results are obtained on X-ray data for Abell 2163 with the computation of a temperature map.}

\section*{Introduction} 

	Within the context of the study of the dynamical status of clusters of galaxies, which is mainly driven by gravitation, analyzing X-ray emission of the hot intra-cluster gas provides key information since this gas is the main component of the baryonic matter inside clusters. Besides accurate information about the cluster dynamical state can be obtained from the gas temperature. 

On the one hand, X-ray data often suffer from a poor statistic leading to low signal-to-noise ratio and a binning process over the detector pixels is most of the time needed, which decreases the spatial resolution. But on the other hand, keeping faint small-scale structures such as cooling flows or compression regions is required in order to get most information on the dynamical state of the clusters, when reconstructing some images and computing some temperature maps. So, we have to look for the minimal loose of resolution for a given signal-to-noise ratio. That is the reason why we introduce a wavelet multiscale restoration algorithm especially suited for few photons counts. We compare it to other adaptive techniques in section \ref{valid_algo} and show our preliminary results about the denoising of XMM data on Abell cluster of galaxies 2163 in section \ref{prelimi_res}.

\section{The denoising algorithm}

\subsection{Why do we use the wavelet transform?}

On the one hand, a Poisson noise process has no proper variation scale due to its non-correlated nature. Its only specific scale is the one related to the detector sampling. On the other hand the signal is a combination of elementary features at different spatial scales. So a multiscale approach is well suited for analyzing the signal providing that the noise contribution can be computed analytically at each scale.

Let us recall that the wavelet transform of a 2D signal $I(x,y)$ is a set of wavelet coefficients $W_{I}(S,k_x, k_y)$ at each scale S. The scale axis can be regularly sampled on a logarithmic scale with base two while preserving all the information. So we use the discrete wavelet transform associated with scales $2^j$ and coefficients $W_{I}(2^j,k_x, k_y)$. Within this dyadic scheme, the wavelet coefficients $W_{I}(2^j,k_x, k_y) \equiv W_{F(2^{j-1})}$ can be computed at each scale $2^j$ from the finer approximation $F_I(2^{j-1},x,y)$ of the signal at scale $2^{j-1}$.

The main idea of our multiscale denoising algorithm is to separate the noise from the signal by {\bf selecting} at each scale the wavelet coefficients with values having a low enough probability to be due to the noise.

\subsection{Denoising with Poisson noise}

Given the statistical characteristics of any noise process, we can estimate whether a given wavelet coefficient value originates from the signal or comes from a chance fluctuation of the involved noise process. To access such information we have to compute at each scale the {\bf Probability Density Function (PDF)} of the wavelet coefficients for the noise process, $p_W (W_{I}(2^j,k_x, k_y))$ . Then a significance threshold $\epsilon$ can be chosen enabling to select coefficients with values higher than $t$ and with a risk to be a chance fluctuation lower than $\epsilon$:
\begin{equation}
	t(2^j,k_x,k_y)=min \left\{ x~\left|~\left[\int_{x}^{\infty} p_W(u) du < \epsilon\right]\right.\right\}.
\end{equation}

Such a selection of the \textit{significant} wavelet coefficients has become a standard technique in case of a Gaussian noise. Given the standard deviation of the Gaussian noise measured in the image at the highest resolution, the PDF can be computed at each scale. The problem with a Poisson statistic is that the noise is signal dependent and so does the PDF. Here, an analytical computation of the PDF at each scale and for each value of the signal at the same resolution is enabled thanks to the use of the unnormalised \textit {Haar wavelet transform}. Hence we can take benefit of a unique and fast procedure, working for both low and high counts of photons.

\subsection{The algorithm}

The denoising restoration algorithm makes use of the unnormalized Haar wavelet transform, which enables to compute thresholds related to the Poisson noise analytically. Within a multiscale approach, this peculiar analyzing wavelet has the unique property to have a shape similar to the square pixel of the detector which provides the image at the highest available resolution. It results also in a decomposition onto three detail spaces where the wavelet coefficients $W_{a,i}$ with $a = h,v,d$ measure the horizontal, vertical and diagonal details respectively.

\begin{figure}
\begin{center}
\includegraphics[height=3.5in, width=3.3in]{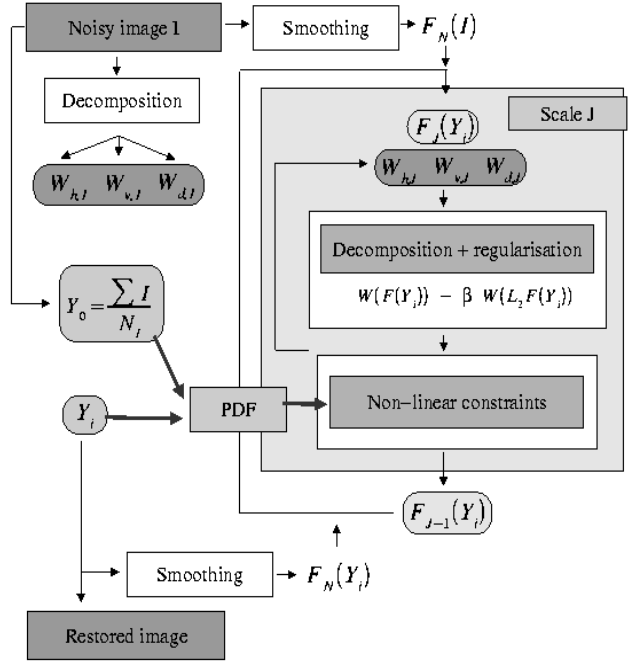}
\caption{Algorithm diagram\label{fig:algo}}
\end{center}
\end{figure}

The set of PDFs is signal-dependent, but the signal is initially unknown. So we have to use an iterative algorithm. Starting from a uniform distribution, we iteratively improve the model $Y_i$ (cf. figure \ref{fig:algo}) of the required signal. A measure of the noise on the wavelet coefficients is computed at each iteration for each scale, using the set of PDFs we are able to compute for each value of the signal at these scales. Using these PDFs we can then define two domains in the wavelet space by selecting the coefficients according to their value, the significant {$W_{a,I}^s(2^j,k_x, k_y)$} and non significant {$W_{a,I}^{ns}(2^j,k_x, k_y)$} ones:

\begin{eqnarray}
	\nonumber W_{a,I}^s\left(2^j,k_x, k_y\right) &=& where \left(\left|W_{a,I}\left(2^j,k_x, k_y\right)\right| > t\left(2^j,k_x,k_y\right)\right)\\
	W_{a,I}^{ns}\left(2^j,k_x, k_y\right) &=& where \left(\left|W_{a,I}\left(2^j,k_x, k_y\right)\right| < t\left(2^j,k_x,k_y\right)\right).\label{equ:signif_domain}
\end{eqnarray}

This selection is done up to the convergence of the model at iteration N, the solution $Y_N$ being the final restored image. This improvement of the model $Y_i$ is done in the wavelet space from the lowest resolution to the highest one, providing a finer and finer image of $Y_{i+1}$ in the direct space. At each step, non-linear constraints on the wavelet coefficients are applied with respect to the initial coefficients of the noisy image: the values of the coefficients inside the significant domain must lay inside a \textit{tolerance range} around the value of those of the initial noisy image, whereas the coefficients of the non-significant domain have to remain lower than the threshold on the confidence level.

The drawback of the use of the Haar wavelet, is that such a steep function entails some blocking effects on the restored image. That's why a regularization step has to be added. So, we introduced an a priori on the expected signal at each scale, using a Tikhonov regularity constraint, leading to gradients as small as possible. To do obtain $L_2 F_Y(2^j) = 0$, we use an iterative Van-Cittert algorithm leading to:
\begin{equation}
	W_{a,F^{(n+1)}(2^j)} = W_{a,F^{(n)}(2^j)} - \beta W_{a,L_2 F^{(n)}(2^j)},
\end{equation}
where $\beta$ is a fixed convergence parameter.

The convergence criterion for the improvement of $Y_i$ is a maximal value for the variation of the total flux of the image, whereas the criterion for the improvement of the wavelet coefficients at each scale is more severe: a maximal value for the variation of the values of each pixel of the image. Such a process with two loops enables a fast convergence on $Y$, --- usually four iterations are enough ---, and few PDF computations are required.

\section{Validation of the algorithm\label{valid_algo}}

\subsection{A toy-model}

\begin{figure}
\begin{center}
\begin{tabular}{cc}
\includegraphics[height=2.in]{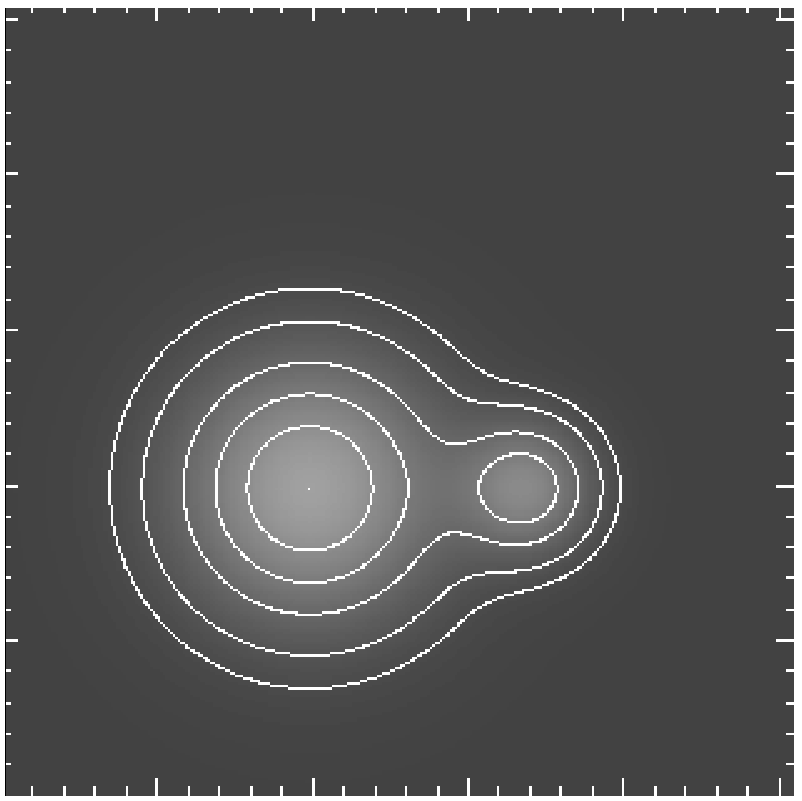} & 
\includegraphics[height=2.in]{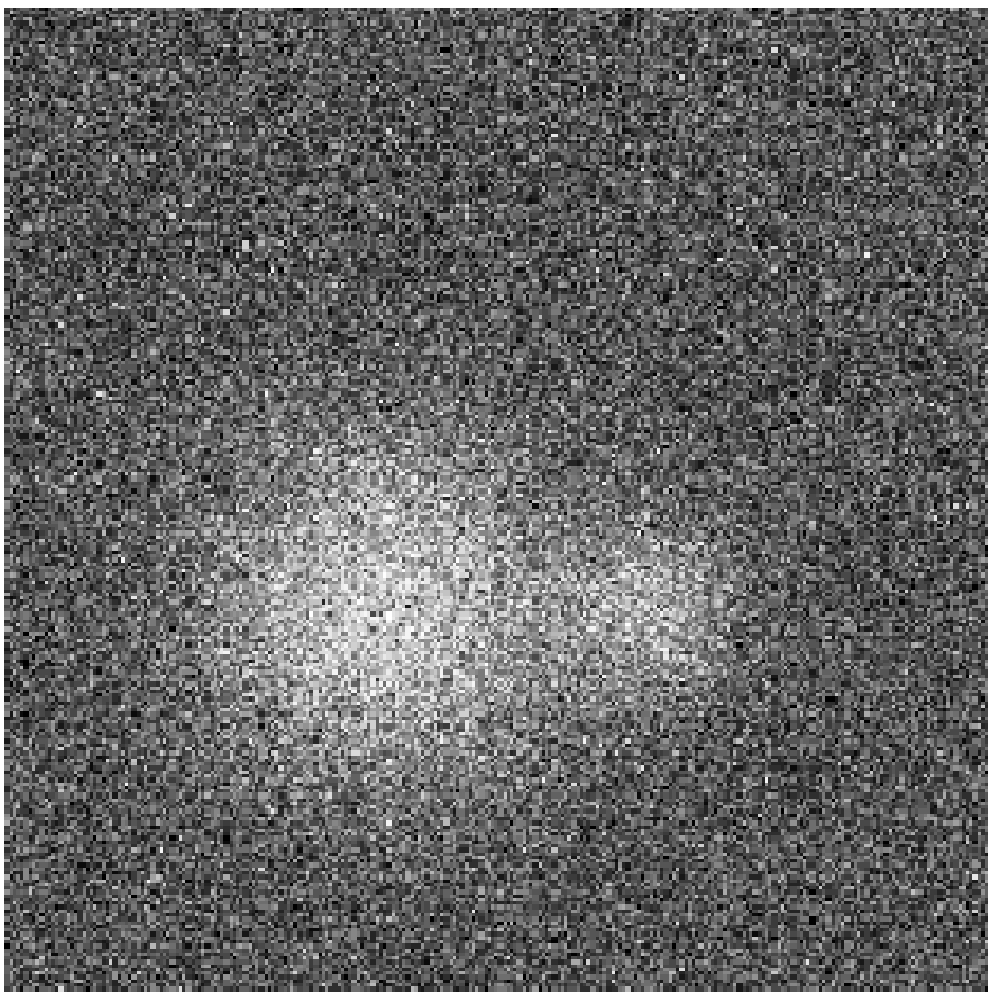} \\
\includegraphics[height=2.in]{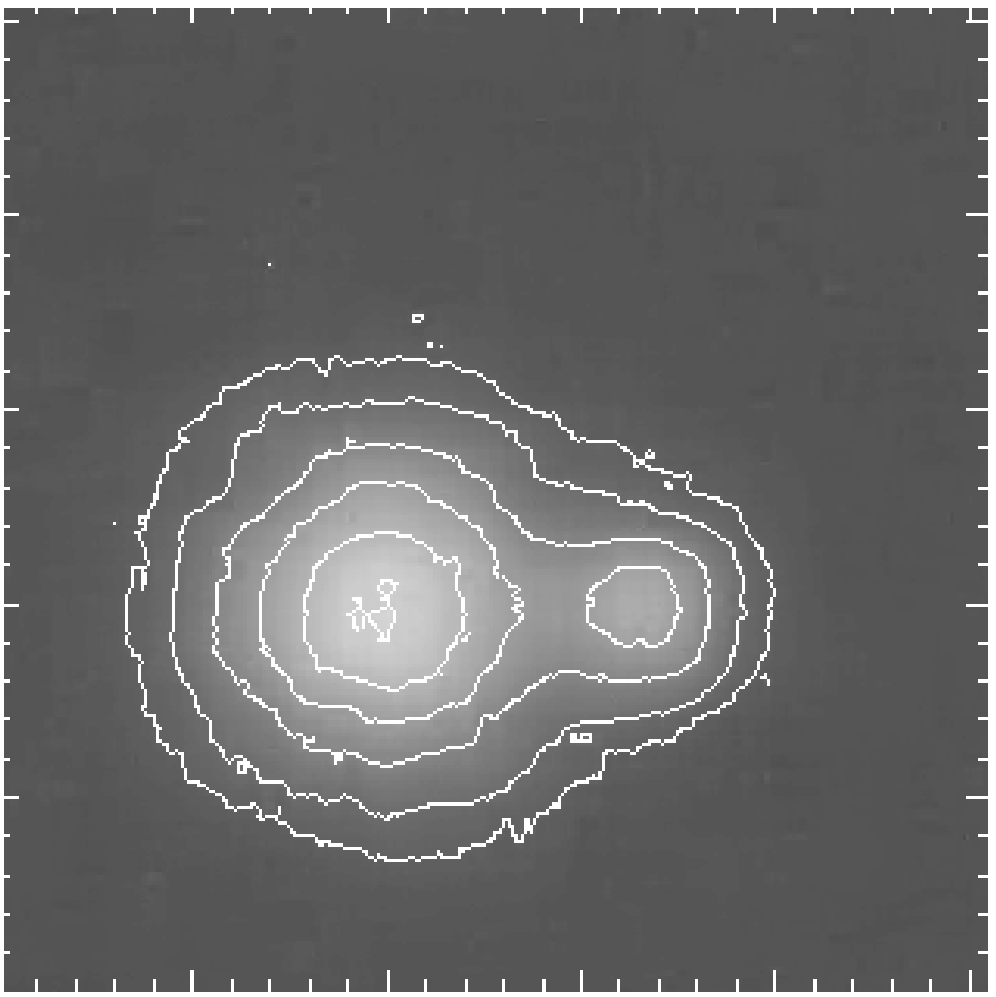} &  
\includegraphics[height=2.in]{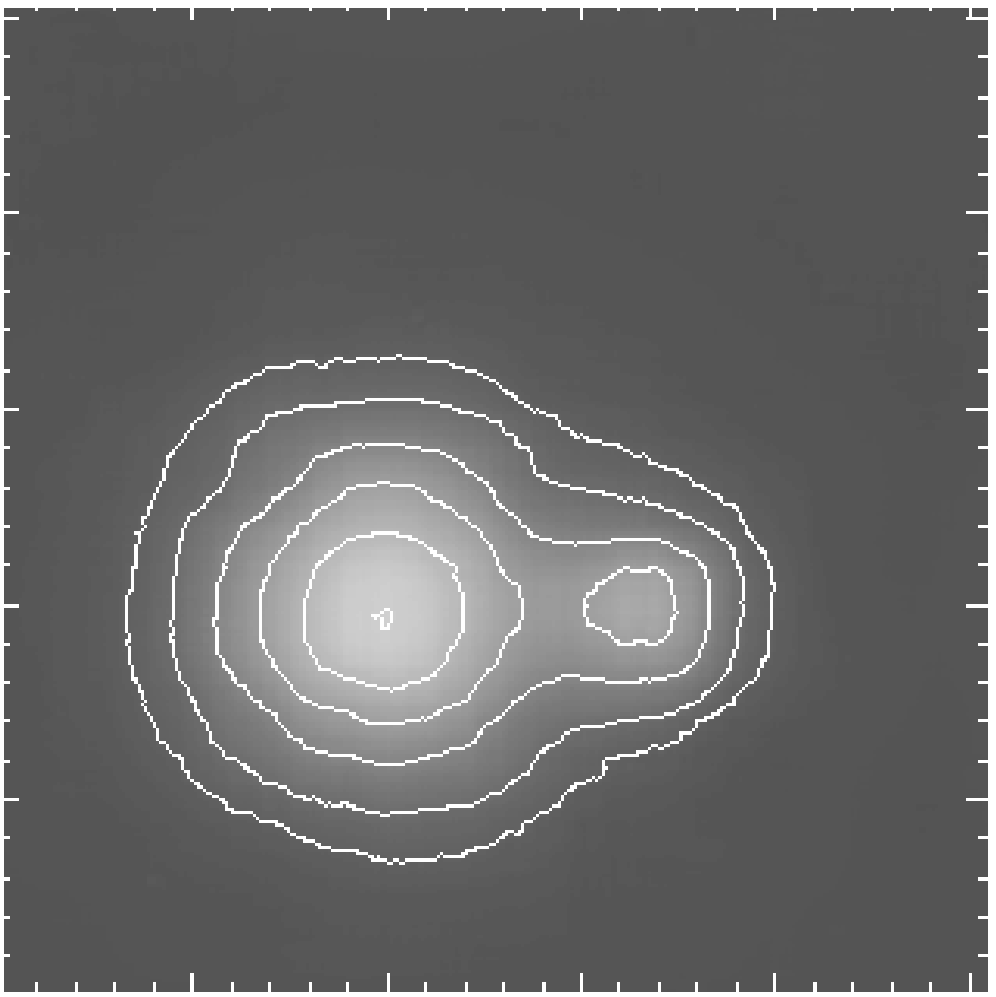} \\
\end{tabular}
\end{center}
\caption{Test on a two Gaussian distributions model. Up-Left: model. $A_1$ = 10, $A_2$ = 7, $\sigma_1$ = 30, $\sigma_2$ = 15, $bck$ = 10. Up-right: noisy image. Down-left: denoising with $10^{-3}$ confidence level. Down-right: denoising with $10^{-5}$ confidence level.\label{fig:toy-model}}
\end{figure}

We first tested the algorithm on simple toy-models, i.e. noisy images of a known parametric model, and verified its ability to restore the initial model, especially the local shape of the image and its total flux. The two critical parameters of the algorithm are the \textit{confidence level} on the probability for a coefficient to be due to the noise, and the \textit {tolerance parameter} which sets the maximum range in which the significant coefficient can be different from their initial values. Figure \ref{fig:toy-model} describes the result for a toy-model of two Gaussian distributions with amplitudes of 10 and 7 counts respectively, plus a background of 10 counts. Whatever the chosen parameter are we reconstruct the model with a total flux discrepancy fewer than a percent, so the restoration technique is robust. But the local shape of the signal (as checked by the isophotes) is better reconstructed using a lower confidence level, especially inside the valley between the two Gaussian distributions. The confidence level must be less severe if we want to detect fainter structures, but obviously the price is that we are less \textit {confident} on their probability to be due to the signal and not to the noise.

\subsection{Wavelet denoising vs. other adaptive methods}

We also compared our new image restoration algorithm with respect to other adaptive techniques like those described in Sanders \& Fabian (2000)\cite{Fabian}. An adaptive binning technique bins the counts using a square kernel, the size of which depends on the signal, so as to get a signal to-noise ratio value higher than a given threshold. An adaptive smoothing approach is a similar technique but using a Gaussian kernel whose width also depends on the signal. 

\begin{figure}[ht]
\begin{center}
\begin{tabular}{cc}
\includegraphics[height=2.in]{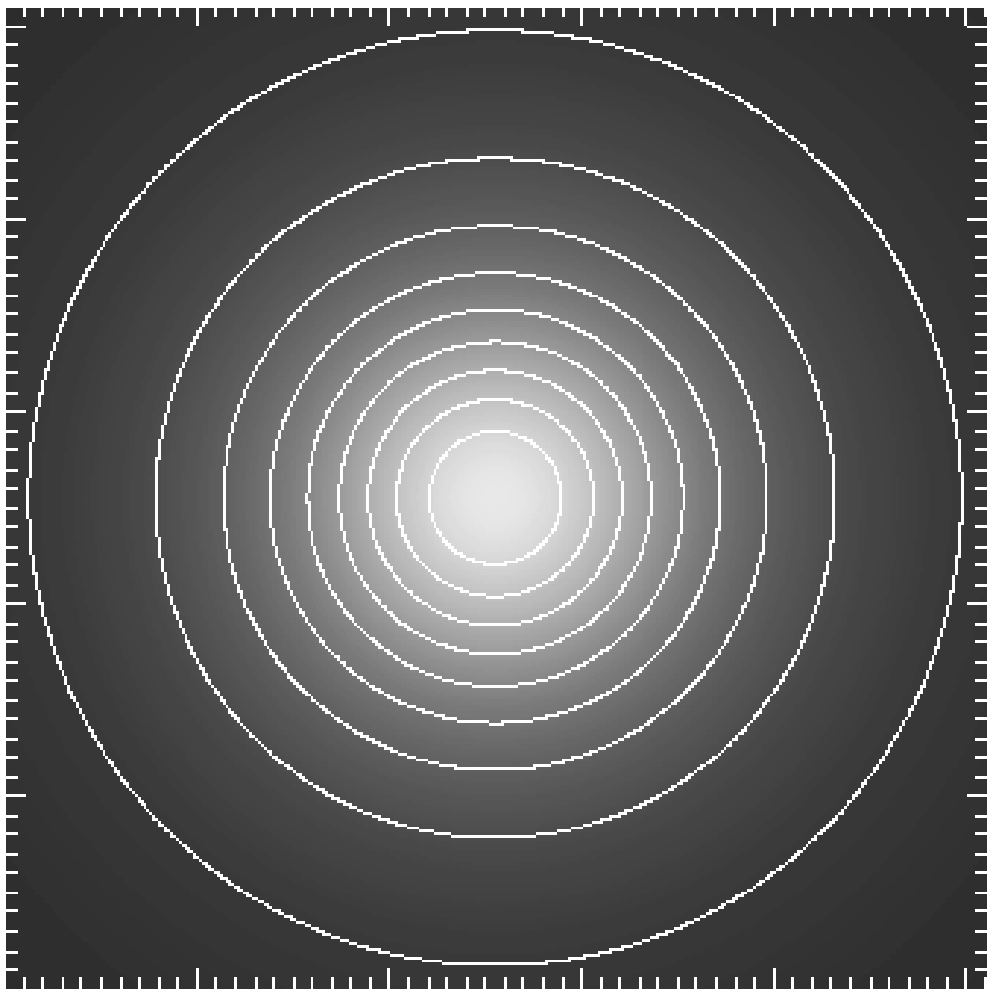} & 
\includegraphics[height=2.in]{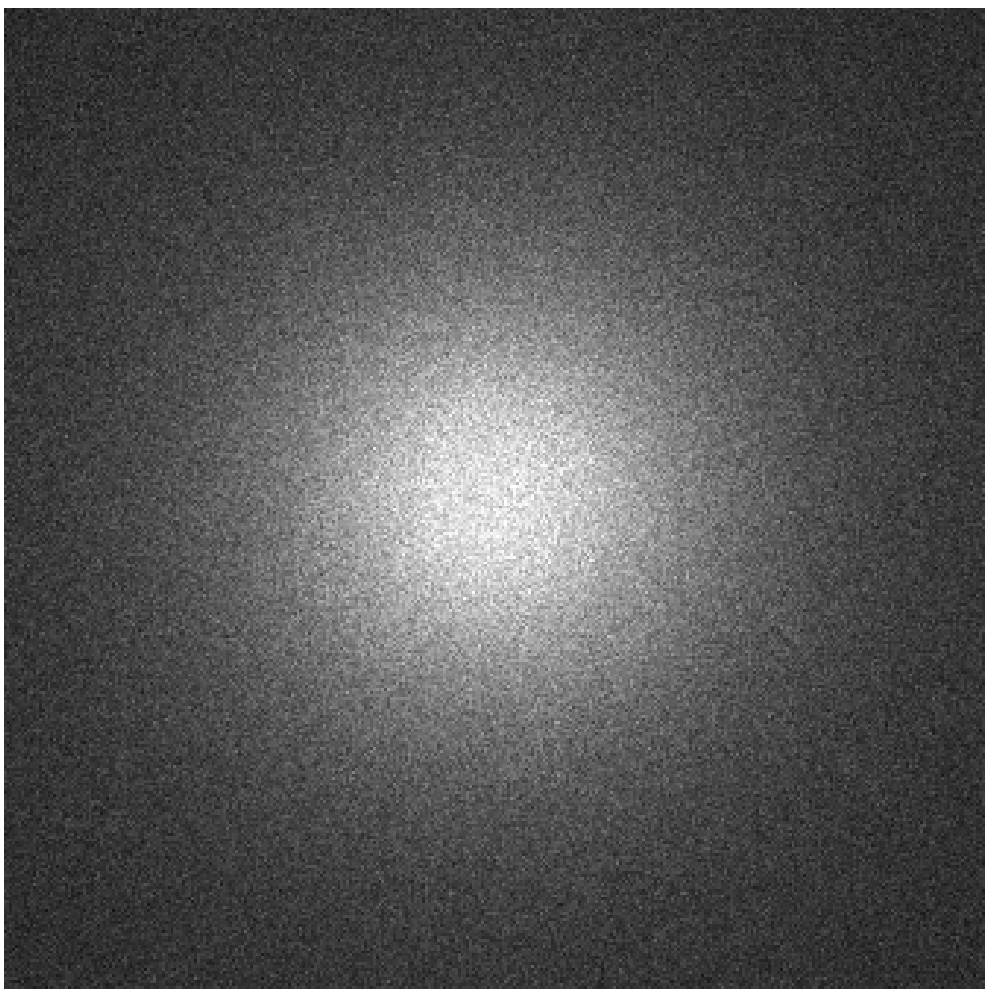} \\
\end{tabular}
\begin{tabular}{ccc}
\includegraphics[height=2.in]{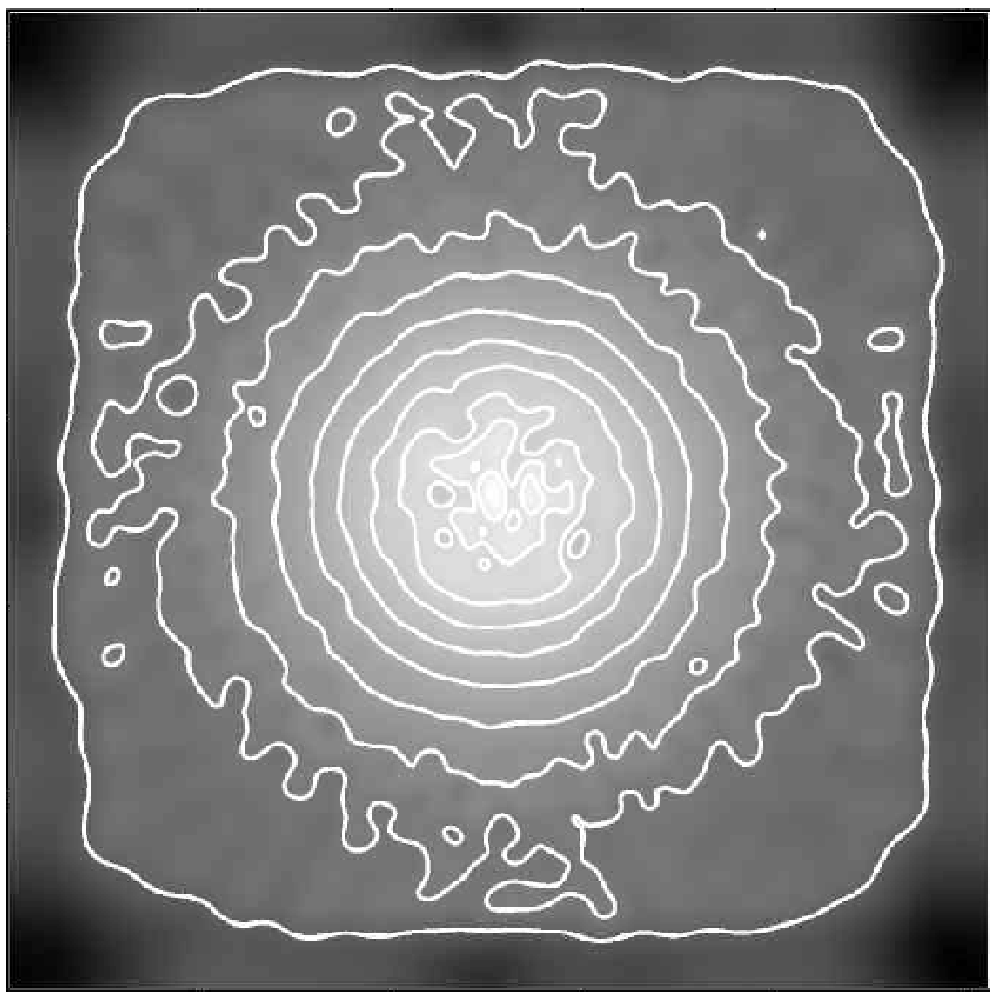} &  
\includegraphics[height=2.in]{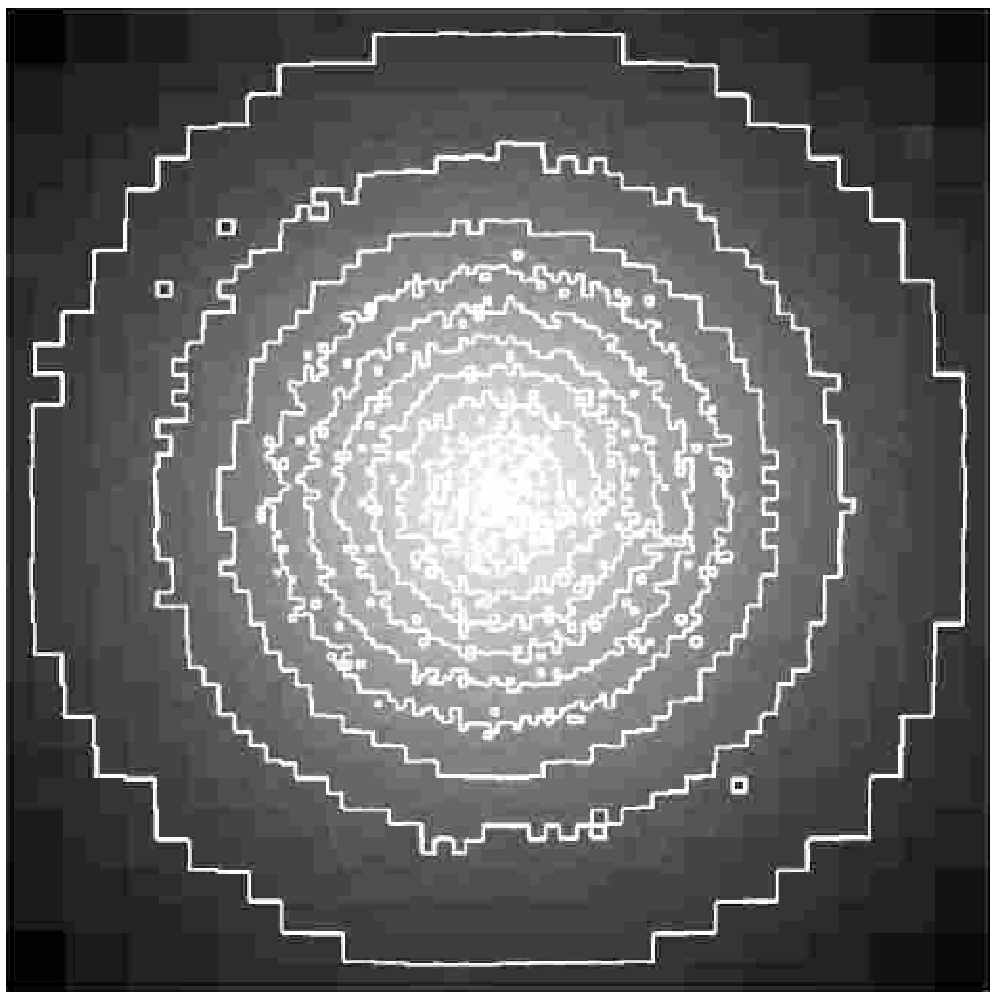} & 
\includegraphics[height=2.in]{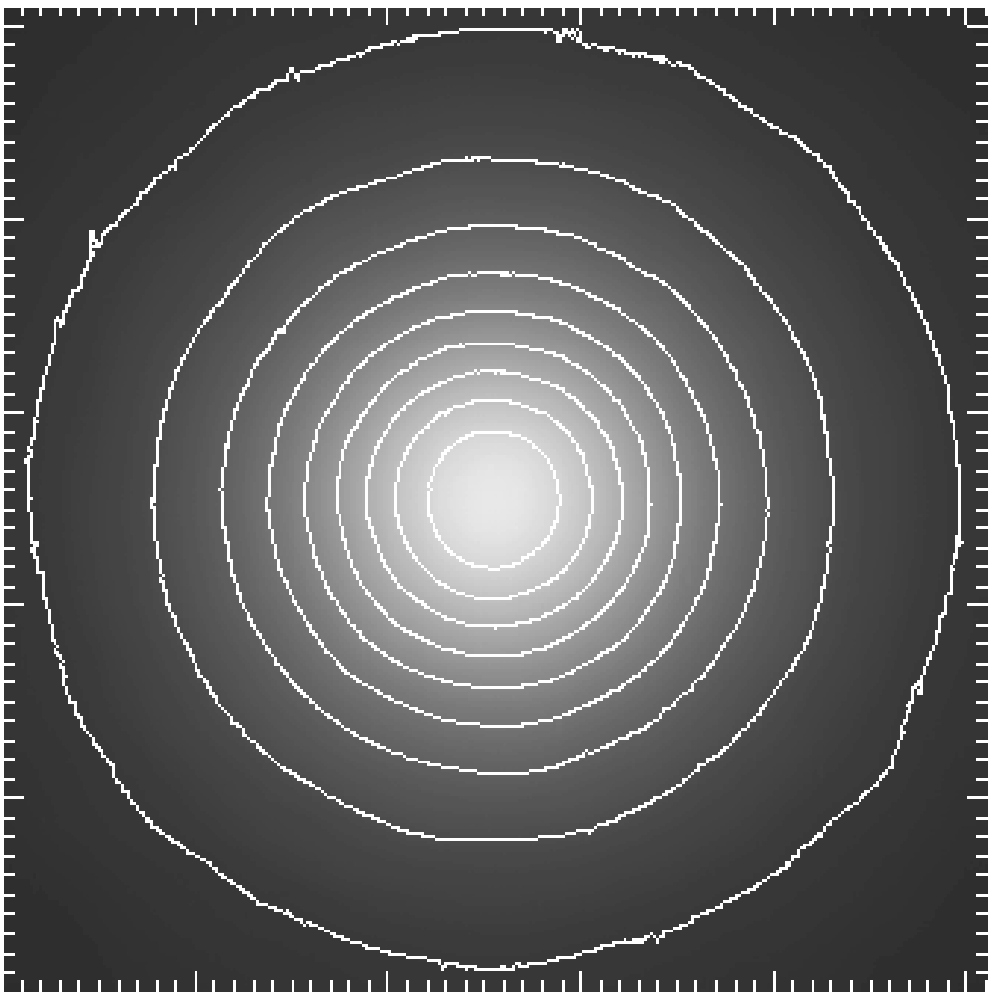} \\
\end{tabular}
\end{center}
\caption{Wavelet denoising and other adaptive techniques. Up-Left: $\beta$ - model, $\beta$ = 0.67, $S_o$ = 100 cts/pix, $bck$ = 20 cts/pix. Up-Right: Poisson noisy image. Down-Left: Adaptive smoothing, $\sigma_{min}$ = 4. Down-Centre: Adaptive binning, error threshold: 0.06. Down-Right: Wavelet denoising. \label{fig:wave_adapt}}
\end{figure}

Let us now consider a $\beta$-model superimposed onto a low background (cf. figure \ref{fig:wave_adapt}). The restoration with adaptive binning and smoothing appears to be good in low signal regions, like the wings of the $\beta$-model, and in regions where the kernel is large, i.e. regions where the local background can be considered as uniform. But in regions of high signal and then of narrow kernels, large noise fluctuations can be misleading and this might explain why such restorations are not accurate for the top of the peak. 
Our algorithm based on the Haar analyzing wavelet can be compared with the adaptive binning technique, since both use steep multiscale adaptive kernels. Our restoration is better at the top of the peak, mainly because a regularization constraint is added in our algorithm forcing the smoothness of the solution.   
The wavelet algorithm should also perform better on reconstructing complex structures, since we iterate on models of the real signal, and do not estimate the flux from the raw noisy image. 

\section{Preliminary results on Abell 2163\label{prelimi_res}}

\subsection{Abell 2163: an exceptional cluster}

Abell 2163 is a rich cluster located at a moderate redshift of 0.2. Central galaxies have a very high velocity dispersion with subgrouping and an extended and luminous radio-halo has been detected. Considering the X-ray emission, Abell 2163 is one of the most luminous and hottest cluster of galaxies. For instance Elbaz et al. (1995)\cite {Elbaz}, combining Ginga spectroscopy and ROSAT PSPC imaging, fitted an isothermal model with $ \text{L}_{\text{X}} (2-10 ~~ \text{keV}) = 6.10^{45} \text{ergs.s}^{-1}$ and $\text{kT} = 14.6 ~~ \text{keV}$, whereas most luminous known clusters have mean temperature and luminosities not exceeding 9 keV and $2 \times 10^{45} \text{erg.s}^{-1}$ respectively. They derived a gas mass of $1.43\times10^{15} \text{M}_{\odot}$ and a binding mass of $4.6\times10^{15} \text{M}_{\odot}$. So Abell 2163 would be $2.6$ times more massive than the Coma cluster. It would represent the extreme high end of the cluster temperature and mass functions, which are a strong constraint for hierarchical CDM structure formation models. Besides, the morphology of its X-ray surface brightness is disturbed, with an elliptically shaped distribution varying with distance to the centre and two brightness enhancements both seen in ROSAT PSPC and XMM images. Eventually, considerable spatial temperature variations in the central region of the cluster have been evidenced by Markevitch et al. using ASCA data\cite{Markevitch_noniso} and now CHANDRA ones.

From these exceptional properties, especially its high mean temperature, galaxy velocity dispersion and gas temperature variations, we can infer that Abell 2163 is probably a non-relaxed recent or ongoing merger.

\begin{figure}[ht]
\begin{center}
\begin{tabular}{cc}
\includegraphics[height=2.in]{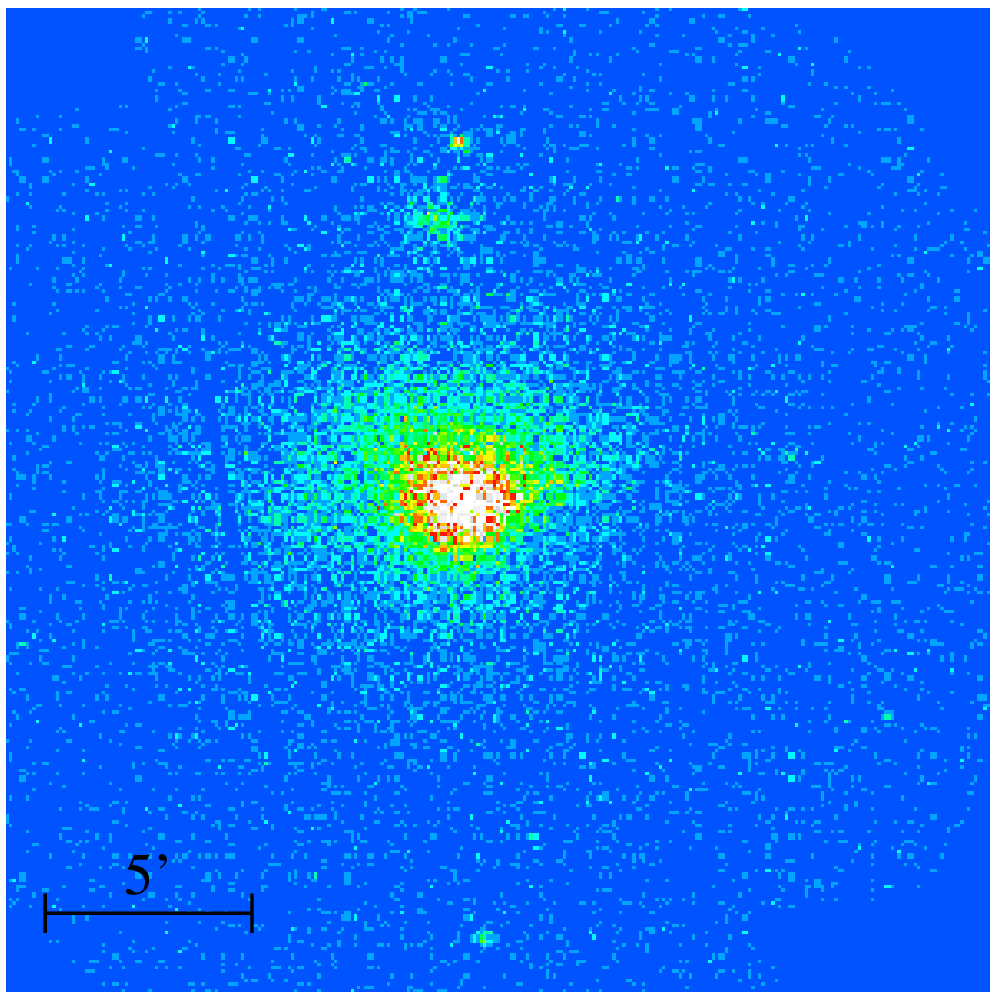} &  
\includegraphics[height=2.in]{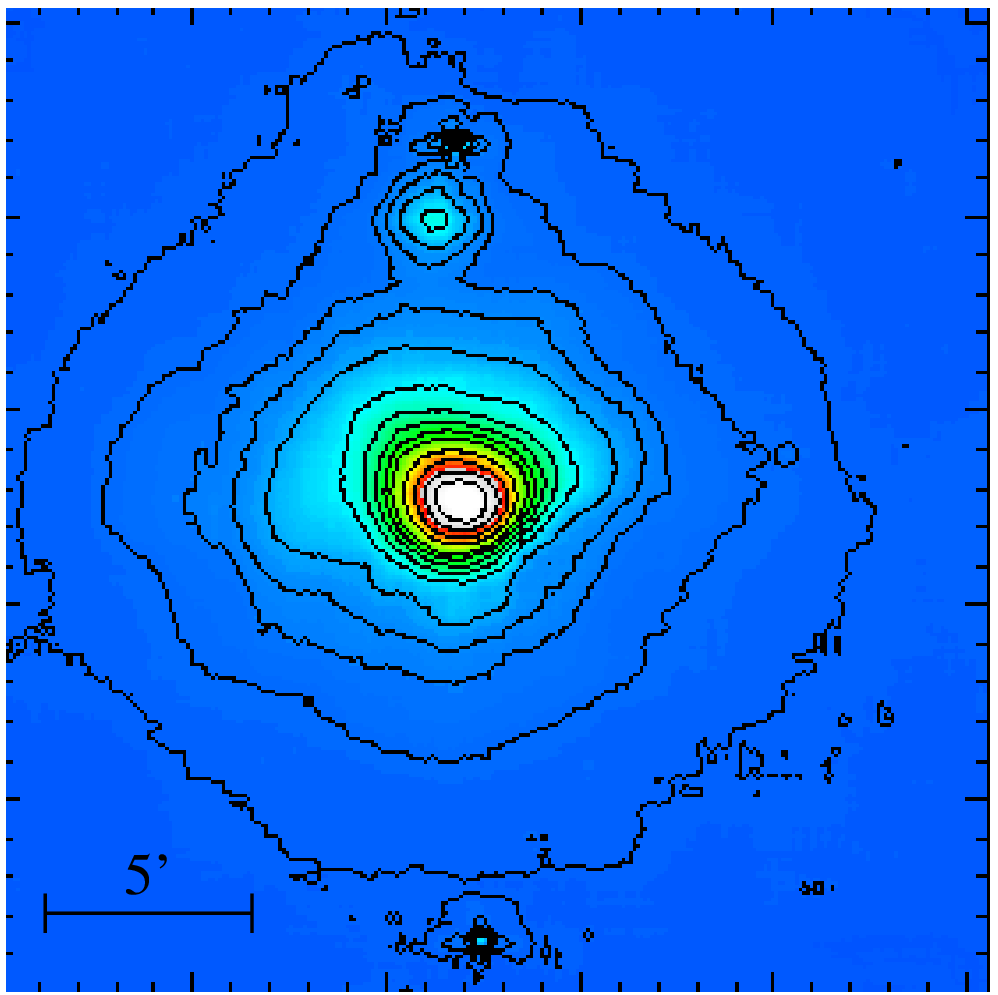} \\
\end{tabular}
\end{center}
\caption{Denoising Abell 2163 map. Left: XMM MOS1, 0.5-2 keV (soft-band). Right: Soft band denoised (0.5-2 keV). \label{fig:denoise_amas}}
\end{figure}

\subsection{Denoising XMM data on Abell 2163}

We used our algorithm for denoising the XMM data on Abell 2163. The raw images come from the analysis of Pratt, Arnaud and Aghanim (these proceedings). We worked with cameras MOS1 and MOS2, composed of seven CCD covering a $0.7$ square degrees field i.e. a $5.1 ~ \text{h}_{100}^{-1} \text{Mpc}^2$ zone at $\text{z} = 0.2$, centered on position: 16h 15m 47s, $-06^{\circ}$ $09^{'}$ $00^{''}$ (equinox 2000.0). Two maps were processed, covering the ``soft band'' in the range 0.5-2 keV and the ``hard band'' in the 2-8 keV one respectively. The exposure time is 2h 55mn, the pixel size is 5.5 arcsecond. XMM data suffer from a uniform particle background. We did no subtraction, since the particle background contribution vs. the cluster contribution is negligible by a factor or more than ten for these energy bands in the central region of the cluster we will focus on.

We applied our denoising algorithm on the XMM mosaic, using six scales and a tolerance parameter of 1 for the four lowest scales and 0.1 for the two highest ones, in order to prevent excessive regularization. We chose a confidence level of $10^{-4}$. The algorithm converged in four iterations. The result can be seen in figure \ref{fig:denoise_amas}. Note that the two external isophotes of the figure are particle background contaminated, so nothing can be said from them. In the central region, we confirm the presence of substructures already seen with ROSAT: a compression region in the South-West, two peaks located in the North and South respectively, and an extended emission region in the North. 

\begin{figure}[ht]
\begin{center}
\begin{tabular}{cc}
\includegraphics[height=2.in]{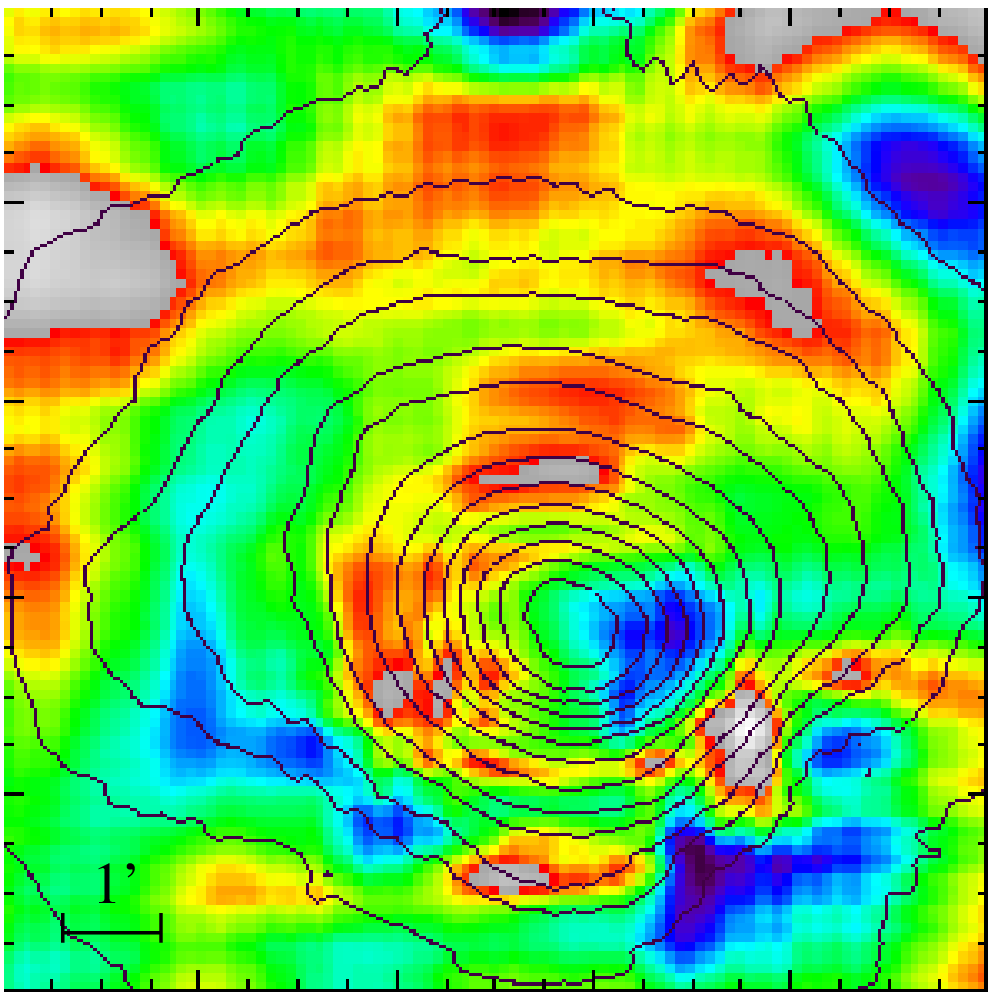} & 
\includegraphics[height=2.in]{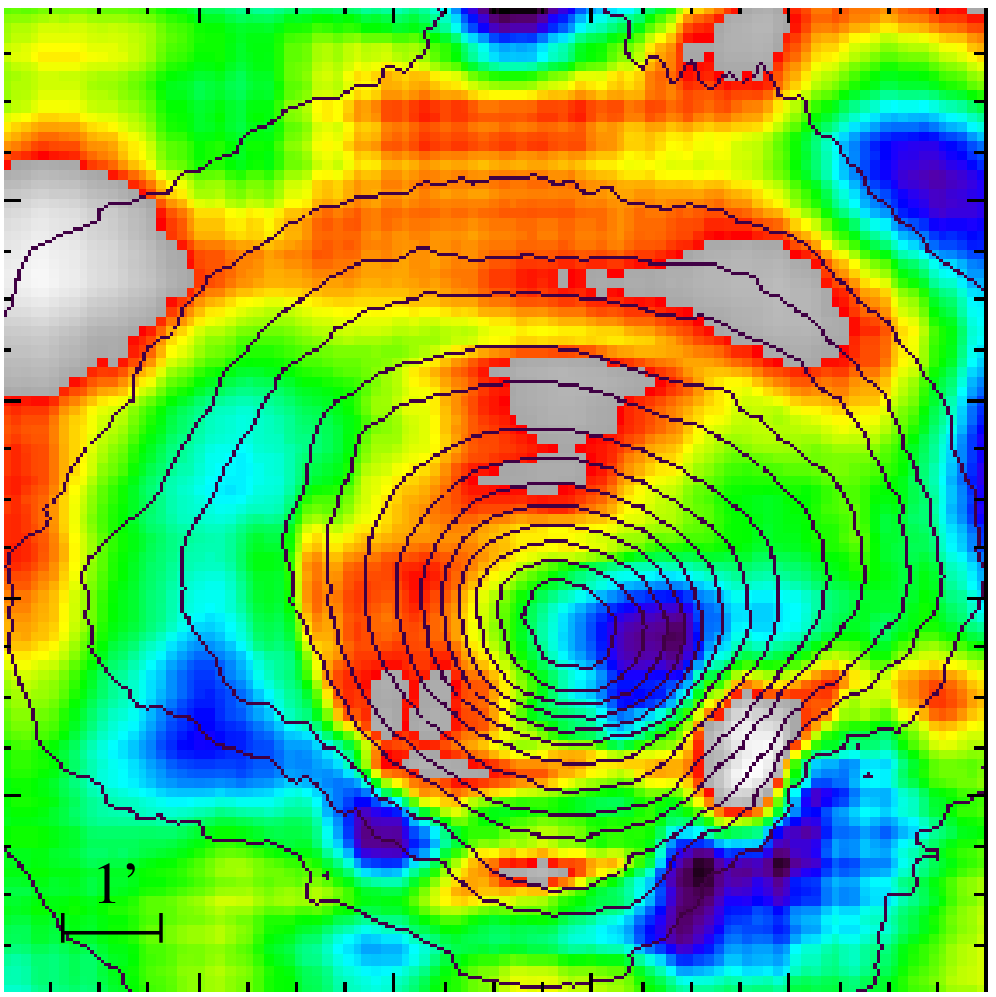} \\
\end{tabular}
\end{center}
\caption{Hardness ratio map of the central region of Abell 2163. Left: threshold, $10^{-2}$. Right: Threshold, $10^{-4}$\label{fig:HR_map}. Hot regions in red. Cold regions in blue.}
\end{figure}

\subsection{A coarse temperature map of the central region of Abell 2163}

As a preliminary temperature map of the central region of Abell 2163, we computed a hardness ratio map between the previous defined soft and hard bands. We restricted our study to the central region where the particle background is negligible and used the significant regions in the wavelet space which are in common between the hard and soft reconstructed maps. This method, which still needs to be properly validated, enables a significant reduction of the artifacts coming from a crude subtraction of maps, especially in regions of similar flux and shape.

Whatever the chosen confidence level is ($10^{-4}$ or $10^{-2}$), we evidenced a hot North-East arc-like feature surrounding the core, and a cold emission near the compression region. Then, considering the $10^{-2}$ confidence level, this cold region can be divided into three clumps, whereas it appears as a unique feature at the $10^{-4}$ level. However we can not conclude here about the presence of substructures appearing at such a low confidence level. Individual spectroscopy of the supposed clumps is required in order to confirm their existence, providing that no further information can be gained from the X-ray surface brightness study.

\section{Conclusion}

We have introduced a new multiscale denoising algorithm especially suited for images with few photons counts. The separation between the noise and the signal contribution in the data is done iteratively in the wavelet space. The Probability density Function of the wavelet coefficients for Poisson noise are computed analytically at each scale, using the same procedure whatever the count value is. Besides a regularization step is invoked so as to remove residuals due to the use of the Haar analyzing wavelet, and to the large excursions specific to Poisson process with a low mean value.

Considering tests on toy-models of extended sources with background, the algorithm is robust at recovering the total flux and the local shape of the signal. The regularization step seems to improve significantly the quality of the restoration. Introducing such extra information is reasonable for X-ray surface-brightness maps where no sharp edges are expected.

We used this algorithm for denoising XMM counts on the Abell cluster of galaxies 2163. Its complex structure is well reconstructed thanks to the use of our multiscale approach. We also computed a hardness ratio map of the centre of the cluster, evidencing a hot arc-like structure around the cluster centre, and a cold emission region corresponding to a South-West compression area. These results confirm the Markevitch et al.\cite{Markevitch_Moriond} conclusions obtained using Chandra data and a spectral fitting on five energy bands.

These promising results confirm that our algorithm should be useful to get accurate surface brightness maps from XMM data. A similar approach taking into account the spectral information should also provide accurate temperature map, which is especially welcomed for studying merging clusters.

\section*{Acknowledgments}
We thank G. Pratt and N. Aghanim for allowing us to use the XMM images of A2163 and for useful discussions on the restored images.


\end{document}